\documentclass[prb,nofootinbib,twocolumn,superscriptaddress]{revtex4} 

\usepackage{graphicx}
\usepackage{dcolumn}
\usepackage{bm}
\usepackage{threeparttable}
\usepackage{times}
\usepackage{mathptmx}
\usepackage{lscape}
\usepackage{natbib}
\usepackage{amsmath}
\usepackage{amssymb}
\usepackage{braket}
\usepackage{comment}
\usepackage{color}

\def\degree{\kern-.2em\r{}\kern-.3em}

\begin{document}


\title{  Formulation of Genuine Thermodynamic Variables from Special Microscopic States   }

\author{Koretaka Yuge}
\affiliation{
Department of Materials Science and Engineering,  Kyoto University, Sakyo, Kyoto 606-8501, Japan\\
}%

\author{Shouno Ohta}
\affiliation{
Department of Materials Science and Engineering,  Kyoto University, Sakyo, Kyoto 606-8501, Japan\\
}%

\begin{abstract}
{
For classical discrete systems under constant composition, it has been considered that genuine thermodynamic variables such as free energy cannot be generally determined from information about a single or a few selected microscopic states. Despite this fact, we here show that Helmholtz free energy for any given composition for disordered states can be well characterized by information about a few ($R$+3, where $R$ denotes number of components) specially selected microscopic states, whose structure can be known \textit{a priori} without requiring any thermodynamic information. The present study is a non-trivial extension of our recently-developed theoretical approach for special microscopic states in canonical ensemble to semi-grand canonical ensemble, which additionally enables to characterize temperature dependence of other thermodynamic variables such as internal energy and entropy. 
  }
\end{abstract}


\maketitle

\section{Introduction}
In classical discrete systems, Helmholtz free energy is typically given by
\begin{eqnarray}
F = -\beta^{-1} \ln Z,
\end{eqnarray}
where $\beta$ denotes inverse temperature $\left(k_{\textrm{B}}T\right)^{-1}$ and $Z$ denotes partition function given by
\begin{eqnarray}
Z = \sum_I \exp\left(-\beta E_I\right),
\end{eqnarray}
where summation is taken over all possible microscopic states on configuration space.
Since number of possible states should astronomically increase with increase of system size, it is generally impossible to exactly determine the value of 
free energy for practical systems.
Therefore, several theoretical techniques have been developed to effectively sample important microscopic states, such as thermodynamic integration\cite{ti1,ti2,ti3,ti4} and Wang-Landau algorism.\cite{wl} 
The reason for performing such numerical simulation comes from the fact that a set of microscopic states having dominant contribution to $F$ should in principle depends on temperature and system (i.e., potential energy landscape), clearly stated by classical statistical mechanics. 

Despite these facts, we recently find a special microscopic state, called projection state, which can characterize macroscopic structure in thermodynamically equilibrium state, where its structure is independent of temperature and of interactions.\cite{em1,em3,cm,em2} This strongly indicates the importance of geometry on configuration space, i.e., the landscape of configurational density of states (CDOS) for non-interacting system, which can break the curse of dimensionality lies under the current classical statistical mechanics.
The present study perform non-trivial extention of our previous work to semi-grand canonical ensemble, in order to figure out a set of special microscopic states that always exhibit dominant contribution to genuine thermodynamic variables such as Helmholtz free energy. The details are shown below. 

\section{Derivation and Concept}
Since marginal distribution of a fixed composition can be well characterized by multidimensional gaussian and additional odd (especially, the third) order genelarized moments, we can reasonally start from expanding semi-grand canonical average of composition $q_1= 2x-1$ in terms of the generalized moments, namely, 
\begin{eqnarray}
\label{eq:sg}
&&\Braket{q_1}_{\textrm{S}}\left(\beta\right) = \Braket{q_1} -\beta \sum_{\alpha=1}^f \Braket{q_1 q_\alpha}\Braket{I|q_1}\Braket{I|q_\alpha} \nonumber \\
&&+ \frac{\beta^2}{2} \sum_{\alpha,\beta=1}^f \Braket{q_1 q_\alpha q_\beta}\Braket{I|q_1}\Braket{I|q_\alpha}\Braket{I|q_\beta} + \cdots,
\end{eqnarray}
where
\begin{eqnarray}
I = E-\Delta \mu N \frac{\left(q_1 + 1\right)}{2}
\end{eqnarray}
and $\Delta \mu$ denotes difference in chemical potential, $\Delta \mu = \mu_{\textrm{A}} - \mu_{\rm{B}}$.
Here, $\Braket{\quad}$ represents semi-grand canonical average for \textit{non-interacting} system, $\Braket{\quad|\quad}$ denotes inner product on configuration space, i.e., trace over 
possible states for whole composition, and subscript for $q$ denotes dimension of the figure.
Based on the symmetric definition of spin variable $\sigma \pm 1$, we can vanish the exactly zero-contribution to the semi-grand canonical average. In a similar fashion to derive special microscopic states for canonical average, we can first rewrite Eq.~(\ref{eq:sg}) up to 2-order tensor contributions, namely
\begin{widetext}
\begin{eqnarray}
\label{eq:qs}
\Braket{q_1}_{\textrm{S}}\left(\beta\right) &\simeq& \Braket{q_1} - \beta \Braket{q_1^2}\Braket{I|q_1} + \frac{\beta^2}{2}\left( 2\cdot\sum_m \Braket{q_1^2 q_{2_m}}\Braket{I|q_1}\Braket{I|q_{2_m}} + 2\cdot\sum_{r\ge2} \sum_{m,n}\Braket{q_1 q_{r_m} q_{\left(r+1\right)_n}} \Braket{I|q_{r_m}}\Braket{I|q_{\left(r+1\right)_n}}   \right) \nonumber \\
&-&\frac{\beta^3}{6}  \sum_{r\ge 2} \sum_{m, n}\sum_{a=0,2} K\cdot\left(\Braket{q_1^2 q_{r_m} q_{\left(r+a\right)_n} } - \Braket{q_1^2}\Braket{q_{r_m} q_{\left(r+a\right)_n}}  -2\Braket{q_1 q_{r_m}}\Braket{q_1 q_{\left(r+a\right)_n}}   \right)     \Braket{I|q_1} \Braket{I|q_{r_m}} \Braket{I|q_{\left(r+a\right)_n}} \nonumber \\
&-&\frac{\beta^3}{6} \sum_s 3\cdot \left( \Braket{q_1^3 q_{3_s}} - 3\Braket{q_1^2}\Braket{q_1q_{3_s}}  \right) \Braket{I|q_1}^2 \Braket{I|q_{3_s}}
\end{eqnarray}
\end{widetext}
where the subsubscripts denote the class of figure in a given dimension. $K=3$ when $a=0$ and $m=n$, and $K=6$ for otherwise.
We here would like to formulate the semi-grand canonical average in terms of energy for special microscopic states whose structure can be known without any thermodynamic information. Since Eq.~(\ref{eq:qs}) can be rewritten by using quadratic forms for the 2-order tensor contributions, we perform singlar value decomposition (SVD):
\begin{widetext}
\begin{eqnarray}
\Braket{q_1}_{\textrm{S}}\left(\beta\right) &\simeq& \Braket{q_1} - \beta \Braket{q_1^2}\Braket{I|q_1} + \frac{\beta^2}{2} \left( 2\Braket{I|q_1}\sum_m \Braket{q_1^2 q_{2_m}}\Braket{I|q_{2_m}} +  \sum_{c=1}^{f'} {}^t\!X \lambda_c \left(\mathbf{U}_c \otimes \mathbf{V}_c\right) X   \right) \nonumber \\
&-& \frac{\beta^3}{6} \cdot 3\Braket{I|q_1} \sum_{d=1}^{f''} {}^t\!Y \eta_d \left(\mathbf{T}_d \otimes \mathbf{W}_d\right) Y  - \frac{\beta^3}{6}\cdot 3\Braket{I|q_1}^2 \sum_s \left( \Braket{q_1^3 q_{3_s}} - 3\Braket{q_1^2}\Braket{q_1q_{3_s}}  \right)  \Braket{I|q_{3_s}}    , 
\end{eqnarray}
\end{widetext}
where $\lambda$ and $\eta$ denote singular values, and $\mathbf{U}_c$, $\mathbf{V}_c$, $\mathbf{T}_d$ and $\mathbf{W}_d$ are respectively $c$-th and $d$-th column of matrix $\mathbf{U}$, $\mathbf{V}$, $\mathbf{T}$ and $\mathbf{W}$ given by the SVD:
\begin{eqnarray}
\mathbf{A} &=& \mathbf{UDV}^T \nonumber \\
\mathbf{B} &=& \mathbf{TD'W}^T \nonumber \\
\end{eqnarray}
and
\begin{eqnarray}
A_{ij} &=& \Braket{q_1 q_i q_j} \quad \nonumber \\ 
&&\left(i\in r_m, j\in r'_n, r\ge2, r'\ge2\right) \nonumber \\
B_{i'j'} &=& \Braket{q_1^2 q_{i'} q_{j'}}  -\Braket{q_1^2}\Braket{q_{i'}q_{j'}} - 2\Braket{q_1q_{i'}}\Braket{q_1q_{j'}} \nonumber \\
&&\left(i'\in r_m, j'\in r'_n, r\ge2, r'\ge2\right).  \nonumber \\
\end{eqnarray}
$X$ and $Y$ are $f'-$ and $f''-$ dimensional vectors whose components consist of a set of $\Braket{I|q_g}$, where figure $g$ is included in individual summations. 
When we define that $\lambda_M$ and $\eta_M$ are the largest singular values for respective SVD, we can express semi-grand canonical average for composition only by energy (and given chemical potential) of "7" special microscopic states:
\begin{eqnarray}
&&\Braket{q_1}_{\textrm{S}}\left(\beta\right) \simeq \Braket{q_1} - \beta \Braket{q_1^2} I_1 + \frac{\beta^2}{2} \left( 2 I_1\cdot I_2 + I_3\cdot I_4 \right) \nonumber \\ 
&&- \frac{\beta^3}{2} \left( I_1\cdot I_5 \cdot I_6 -  I_1^2\cdot I_7\right).
\end{eqnarray}
The corresponding microscopic structures are explicitly given by
\begin{eqnarray}
\label{eq:str}
&&\mathrm{str1:} \left\{1, 0, 0, \cdots , 0\right\} \nonumber \\
&&\mathrm{str2:} \left\{0, \Braket{q_1^2 q_{2_1}}, \cdots, \Braket{q_1^2 q_{2_e}}, 0, \cdots, 0   \right\} \nonumber \\
&&\mathrm{str3:} \left\{0, \sqrt{\lambda_M}U_{1M}, \cdots, \sqrt{\lambda_M}U_{f'M}  \right\} \nonumber \\
&&\mathrm{str4:} \left\{0, \sqrt{\lambda_M}V_{1M}, \cdots, \sqrt{\lambda_M}V_{f'M}  \right\} \nonumber \\
&&\mathrm{str5:} \left\{0, \sqrt{\eta_M}T_{1M}, \cdots, \sqrt{\eta_M}T_{f''M}  \right\} \nonumber \\
&&\mathrm{str6:} \left\{0, \sqrt{\eta_M}W_{1M}, \cdots, \sqrt{\eta_M}W_{f''M}  \right\} \nonumber \\
&&\mathrm{str7:} \left\{0, 0,\cdots, Q_{3_1}, \cdots, Q_{3_{e'}} , 0, \cdots, 0 \right\},
\end{eqnarray}
where
\begin{eqnarray}
Q_{3_s} = \Braket{q_1^3 q_{3_s}} - 3\Braket{q_1^2}\Braket{q_1 q_{3_s}}.
\end{eqnarray}
It is now clear from Eq.~(\ref{eq:str}) that structure of the 7 special microscopic states can be known \textit{a priori} without any thermodynamic information, since the corresponding values can be obtained from information about CDOS for \textit{non-interacting} system. 
Therefore, by using the standard relationship in thermodynamics of 
\begin{eqnarray}
\Delta \mu = \frac{\partial F}{\partial \left(Nx\right)},
\end{eqnarray}
we can quantitatively determine the value of Helmholtz free energy $F$ at any given temperature and composition. 

We finally demonstrate how to estimate the values for special states.
For instance, non-zero contribution of $\Braket{q_1^2\cdot q_{2_i}}$ is given by
\begin{eqnarray}
&&\Braket{q_1^2\cdot q_{2_i}} = \frac{1}{N^2\left(DN\right)}\sum_k F_{112_i}\left(k\right) \Braket{q_k} \nonumber \\
&&= \frac{1}{N^2\left(DN\right)} F_{112_i}\left(0\right) \Braket{q_0} = \frac{1}{N^2\left(DN\right)}\cdot 2DN \cdot 1 = \frac{2}{N^2}, \nonumber \\
\quad
\end{eqnarray}
where the factor 2 in numerator comes from the permutation of two 1-body figure. 
Other non-zero contribution to the 3-order moment should always consist of one 1-body, and $m$- and $m+1$-body figure in order to avoid odd-times occupation of the constituent lattice point. 
The lowest dimension diagram corresponds to the 1-, 2- and 3-body figure, are explicitly given by
\begin{eqnarray}
&&\Braket{q_1\cdot q_{2_i}\cdot q_{3p}} = \frac{1}{N\cdot DN\cdot TN}\sum_k F_{12_{i}3_{p}}\left(k\right) \Braket{q_k} \nonumber \\
&&=\frac{1}{N\cdot DN\cdot TN} F_{12_{i}3_{p}}\left(0\right) \Braket{q_0} = \frac{J}{N^2D},
\end{eqnarray}
where $J=4-N_p$ ($N_p$ is the number of pair $2_i$ included in a single triplet figure $3_p$) if the triplet $3_p$ includes at least one pair $2_i$, and $J=0$ for otherwise. 

\section{Conclusions}
Based on geometry of configurational density of states, we derive 7 special microscopic states that always exhibit dominant contribution to Helmholtz free energy, which is invariant for the choice of constituent elements, temperature and potential energy landscape. Significant information about non-interacting system should be re-emphasized in statistical mechanics to break the curse of dimensionality.

\section{Acknowledgement}
This work is supported by a Grant-in-Aid for Scientific Research on Innovative Areas (18H05453) and a Grant-in-Aid for Scientific Research (16K06704) from the MEXT of Japan, Research Grant from Hitachi Metals$\cdot$Materials Science Foundation, and Advanced Low Carbon Technology Research and Development Program of the Japan Science and Technology Agency (JST).

\end{document}